\documentclass[final,english]{bullsrsl}[2023/05/31]

\usepackage[utf8]{inputenc}
\usepackage[T1]{fontenc}
\usepackage{xcolor}
\usepackage{hyperref}
\usepackage[]{natbib}
\usepackage{graphicx}

\hypersetup{
    colorlinks=true,
    linkcolor=red,  
    citecolor=blue,   
    urlcolor=magenta   
}

\begin{document}
\sloppy
\title{Dancing with the stars: a review on stellar multiplicity}

\author[affil={1,2}]{Thibault}{Merle}
\affiliation[1]{Royal Observatory of Belgium, Avenue Circulaire 3, 1180 Brussels, Belgium}
\affiliation[2]{Institut d'Astronomie et d'Astrophysique, Universit\'e Libre de Bruxelles, CP 226 Boulevard du Triomphe, 1050 Bruxelles, Belgium}
\correspondance{\url{thibault.merle@oma.be}}
\date{1st April 2023}
\maketitle

\begin{abstract}
Stars like company. They are mostly formed in clusters and their lives are often altered by the presence of one or more companions. Interaction processes between components may lead to complex outcomes like Algols, blue stragglers, chemically altered stars, type Ia supernovae, as well as progenitors of gravitational wave sources, to cite a few. Observational astronomy has entered the era of big data, and thanks to large surveys like spatial missions Kepler, TESS, \emph{Gaia}, and ground-based spectroscopic surveys like RAVE, \emph{Gaia}-ESO, APOGEE, LAMOST, GALAH (to name a few) the field is going through a true revolution, as illustrated by the recent detection of stellar black holes and neutron stars as companions of massive but also low-mass stars. In this review, I will present why it is important to care about stellar multiples, what are the main large surveys in which many binaries are harvested, and finally present some features related to the largest catalogue of astrometric, spectroscopic and eclipsing binaries provided by the Non-Single Star catalogue of \emph{Gaia}, which is, to date, the largest homogeneous catalogue of stellar binaries. 
\end{abstract}

\keywords{binary stars, stellar multiplicity, spectroscopic binaries, eclipsing binaries, astrometric binaries}


\section{Introduction}
\begin{figure}
\centering
\includegraphics[width=\linewidth]{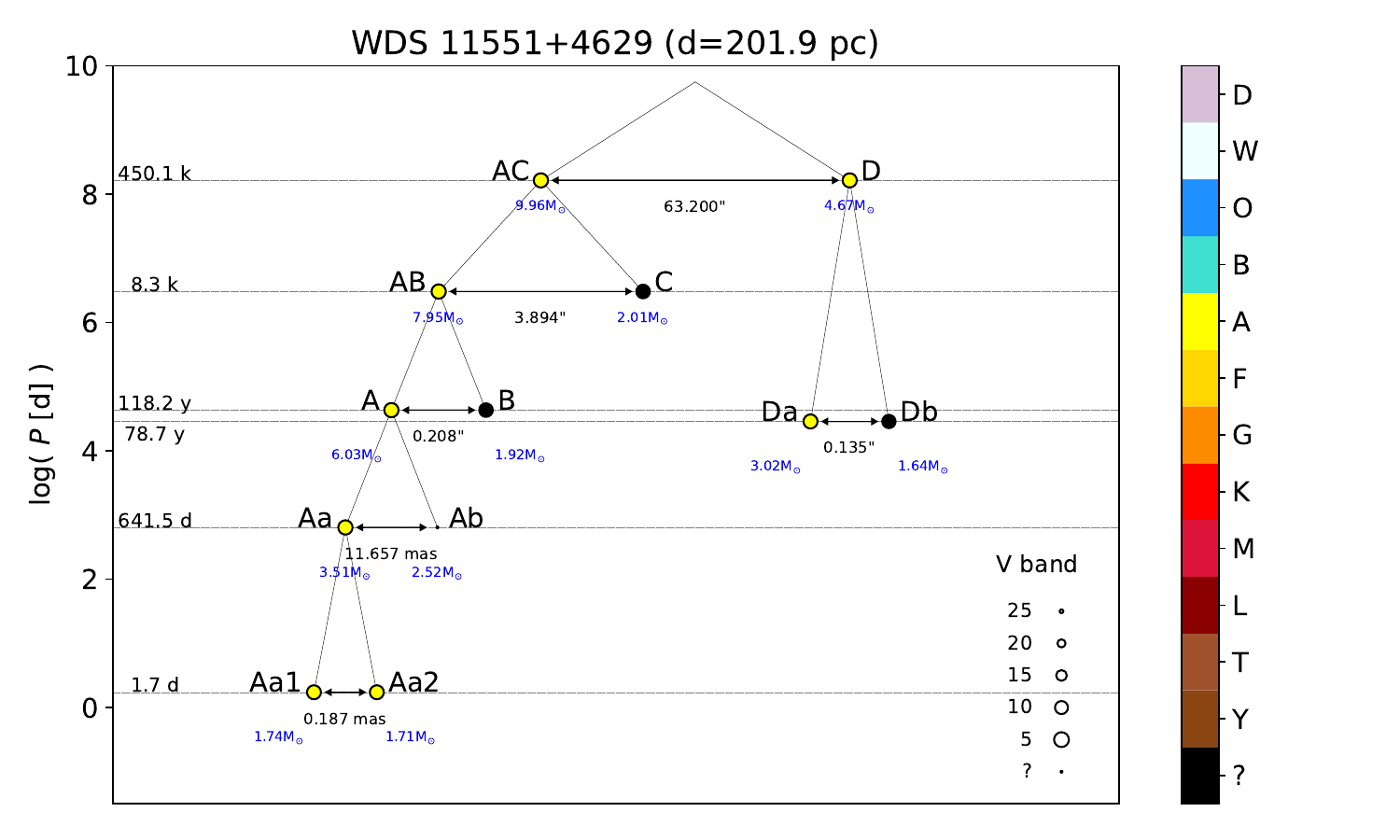}
\caption{Mobile diagram of the stellar septuple 65 UMa with the highest degree of hierachical levels (5) known so far in a 5+2 architecture. The black components  have unknown spectral types. Component C of 2.01 M$_\odot$ could be itself a spectroscopic binary making this system a potential stellar octuple. That is why the follow-up of component C is of very high interest while being challenging (separation of 3.9 arcsec). `k' stands for ky.}
\label{fig:mobile}
\end{figure}

 Most stars form in binaries and hierarchies, in clusters and associations rather than in isolation \citep{duchene2013}. Almost all massive stars have stellar companions \citep{sana2012}: the multiplicity fraction  for O-type main sequence was recently revised to be 94$\pm$14\% while the mean number of companions per early-type star reaches 2.1$\pm$0.3 \citep{moe2017} meaning that most massive stars are part of stellar triples. Nevertheless, the early-type stars (OBA spectral types) represent less than 1\% of all the galactic stars \citep{ledrew2001} and probe only recent local history, impeding galactic archaeology. For long-lived late-type stars, the multiplicity fraction is estimated to be 40 -- 60\%, coming from various samples of the Solar Neighborhood \citep[\emph{e.g.}][]{abt1976, duquennoy1991, raghavan2010, tokovinin2014, fuhrmann2017, moe2017, reyle2021} during the last 50 years (Table~\ref{tab:census}). But the recent analysis of the 10 pc sample, rather complete and based on \emph{Gaia} data \citep{reyle2021}, reduces this fraction to 28\%, meaning a mean number of companions of 0.51, significantly lower than the 0.61 value from \citet{raghavan2010}. At the stellar low-mass end, \citet{winters2019} published  a complete census of the red dwarfs (M spectral type) in the 25 pc sample, with a multiplicity fraction of 27\% marking the lower limit of stellar multiplicity, but in tension with the fraction provided by \citet{moe2017} and \citet{fuhrmann2017}. To be complete, we also mention the recent analyses of wide binaries in the 200 pc and the 1 kpc samples, based on successive \emph{Gaia} releases containing the widest bound systems (up to 0.25 pc, \citealt{el-badry2021}). Stable stellar systems are hierarchical in nature and the highest-order systems known can include 7 components in 5 hierarchical levels like 65~UMa, see Fig.~\ref{fig:mobile}. The Castor system, one of the first physical binary which has been shown to be gravitationaly bound  \citep{herschel1803} is actually a sextuplet made of two binaries orbiting each other with a more distant binary, \emph{i.e.}, a (2+2)+2 architecture, where 3D orbits have been only very recently fully characterised thanks to a combination of spectrometry, photometry and interferometry \citep{torres2022}. 
We warn the reader that the present review is unavoidably biased, and does not pretend to be exhaustive regarding the many aspects of stellar multiplicity. 

\begin{table}
\centering
\caption{Multiplicity statistics in late-type stars. The multiplicity fraction is obtained by summing up the second, third and fourth columns.  The fourth column represents the fraction of quadruples and higher-order stellar systems. 
While the multiplicity fraction is always $\le 1$, the mean number of companions can potentially be larger than one, as it is the case for populations of massive OBA stars, present in negligible proportions in the Solar Neighborhood.
}
\label{tab:census}
\bigskip
\scriptsize
\begin{tabular}{cccccclc}
\hline
\textbf{Singles} & \textbf{Binaries} & \textbf{Triples} & \textbf{Quadruples+} & \textbf{Multiplicity} & \textbf{Mean number }& \textbf{Comments} & \textbf{Ref.} \\
\textbf{[\%]} & \textbf{[\%]} & \textbf{[\%]} & \textbf{[\%]} & \textbf{fraction [\%]} & \textbf{of companions}& & \\
\hline
72.3 & 20.6 & 5.6 & 1.5 & 27.7 & 0.51 & 10 pc sample (339 sys.) & (1) \\
60   & 30   & 9   & 1   & 40   & 0.50 & 25 pc solar-type sample (404 sys.) & (2) \\
47   & 37   & 13  & 5   & 55   & 0.78 & 25 pc solar-type sample (422 sys.) & (3) \\
54   & 33   & 8   & 5   & 46   & 0.64 & 67 pc FG sample (~4850 sys.) & (4) \\
54   & 34   & 9   & 3   & 44   & 0.61 & 25 pc solar-type sample (454 sys.) & (5) \\
57   & 38   & 4   & 1   & 43   & 0.49 & 22 pc FG sample (164 sys.) & (6) \\
42   & 46   & 9   & 2   & 57   & 0.70 & 135 bright FG stars with V < 5.5 & (7) \\
\hline
\end{tabular}
\bigskip

(1) \citet{reyle2021} -- (2) \citet{moe2017} -- (3) \citet{fuhrmann2017} -- (4) \citet{tokovinin2014} -- (5) \citet{raghavan2010} -- (6) \citet{duquennoy1991} -- (7) \citet{abt1976}
\end{table}

\section{Why do we care about stellar multiples?}
Besides the evidence, given in the introduction, on how common binaries and multiples are, stellar multiples are detected through a large variety of observational techniques that probe different period regimes. In this review we will mainly focus on astrometric (AB), spectroscopic (SB) and eclipsing (EB) binaries. We do care about stellar multiples because they allow us to (i) benchmark single stars, understand stellar (ii) formation and (iii) evolution. 

\subsection{Benchmarking}
The mass of a star is the most fundamental parameter for its structure (\emph{e.g.}, convective and radiative zones), its evolution (\emph{i.e.}, dictates which nuclear reactions are taking place), and its final fate as white dwarf (WD), neutron star (NS) or black hole (BH). The stellar objects that provide the most precise and accurate masses (lower than 2\%) are spectroscopic binaries with two visible components that show eclipses (SB2+EB) because they are the least-model dependent \citep{serenelli2021}. They represent cornerstones on which single star evolutionary models are anchored \citep[\emph{e.g.},][]{paczynski1970, iben1984, kippenhahn2013}. They can also be used as benchmarks to calibrate the mass-luminosity, mass-radius and other associated scales for main-sequence stars \citep{eker2018, moya2018}. 

\subsection{Stellar formation}
Our current view of stellar formation relies on elementary formation mechanisms summarised, \emph{e.g.}, in \citet{offner2022}. Stars form by hierarchical collapse of giant molecular clouds caused by the Galactic spiral-arms structure and colliding flows in the interstellar medium \citep{vazquez2019}. The increase of the density of the gas produces a decrease of the Jeans mass that leads to fragmentation in cascade (hierarchical fragmentation, \citealt{bodenheimer1978}) which stops when the gas becomes optically thick and heats adiabatically, increasing the Jeans mass as a consequence. The protostars form at the smallest scales of the cascade. Gravitational dynamics implies that small scales collapse faster than large ones. At each scale, gas infalls from the upper scale, shrinking the orbits of the newly formed systems. \citet{tokovinin2020} showed with simple prescriptions that most close binaries and compact hierarchical triples are indeed formed by disc fragmentation followed by accretion-driven inward migration. In overall, the inside-out formation in which inner pairs form first seems to be the main scenario to explain the formation of triples and 3+1 quadruples although dynamical interactions (captures, disruptions and collisions) can make the outside-in formation possible, for instance to setup the 2+2 quadruples, at least partially \citep{tokovinin2021}.

\subsection{Stellar evolution}
The fate of stellar binaries, as the building blocks of high-order systems, is mainly driven by the masses of the components and their separations. Observationally, binaries are categorised in three groups, following the General Catalogue of Variable Stars \citep{samus2017}: detached, semi-detached and contact binaries. Detached binaries are made of well separated components ($a > 10$~au)  that are supposed to not have interacted in the past (which is not always the case). The semi-detached and contact binaries are generally (very) close binaries ($a < 10$~au) which are currently interacting by means of mass transfer between the components or lost from the system. These binaries are laboratories for studying stellar physics of tidal effects, mass and orbital angular momentum transfers and/or losses \citep[\emph{e.g.}][]{han2020}. Such building blocks are then used to understand the evolution of higher-order systems \citep{Toonen2020}.

Binarity in low-mass stars impacts the individual evolution of the components to create a rich zoo of stellar families \citep[\emph{e.g.}][]{de_marco2017}. Ba stars, CH and carbon-enhanced metal poor (CEMP) stars are classes of chemically-peculiar evolved stars with enhancements in carbon and heavy elements (like barium). These peculiarities originate from binaries that experienced mass transfer in the past, leaving a chemical imprint in the form of an excess of carbon, nitrogen, s-process and/or r-process elements. Such stars are called `extrinsic'. A correlation between orbital period and [s/Fe] was never clearly observed in CEMP stars, contrary to their relatives at higher metallicity (Ba and CH stars, \citealt{jorissen2019}, although not so clear). Metallicity, which modulates the s-process efficiency, may play an important role in that respect \citep{karinkuzhi2021}. In addition, the most metal-poor stars are challenging to characterise because their low metal content produces less lines, and transitions are affected by strong non-LTE effects \citep[\emph{e.g.}][]{ezzeddine2017} that lead to large uncertainties in their astrophysical parameters and abundances. Progresses in determining accurate atomic data are mandatory to obtain precise abundances \citep[\emph{e.g.}][]{merle2011}, especially for iron-peak and heavy elements.

Finally, many astrophysical observations could result from past mergers events occuring in higher-order systems. The supegiant Betelgeuse could be an outcome of a past merger event \citep{chatzopoulos2020} to explain its high spin. The XIXth century  giant eruption in $\eta$~Carinae was suspected to be the result of a merger event in a triple system \citep{portegies_zwart2016}.  R~carbon stars could result from a merger of a red giant with a He white dwarf \citep{mac_clure1997,izzard2007}. Ba stars could arise from pollution of an AGB star to an inner binary that ultimately merges \citep{gao2023}. Most close binaries, including contact binaries, are thought to be formed in triples that undergo Kozai-Lidov oscillations with tidal friction \citep{bataille2018}. Stellar hierarchies that evolved differently from simple binaries can also form more exotic systems like blue stragglers \citep[\emph{e.g.}][]{Geller2013} or type~Ia supernovae through sequence of merger events \citep{merle2022}. On another side, hierarchical systems can cause false positives in the search of exoplanets \citep{santerne2013}.
It is also probable that we can see stellar mergers through transient events like (luminous) red novae, \emph{e.g.}, the 2002 eruption in V838~Mon \citep{kaminski2021} that probably originated from a merger in a triple or higher-order system. Indeed, the increase of luminosity over a short time scale is a sign of accretion through mergers. More dramatic outcomes arise when more massive components interact, as recently revealed by the detection of gravitational waves produced by the merger of neutron stars and black holes \citep{abbott2016,abbott2017}.

\section{Multiple stars in the era of massive surveys}
The observational astronomy has entered the era of big data with (i) large spatial-based photometric surveys mainly devoted to the detection of exoplanets, and (ii) large ground-based spectroscopic surveys mainly devoted to the Galactic archaeology. While such surveys were not generally thought to optimise binary detection, such large and homogeneous samples have triggered systematic hunting of binaries and higher multiples to produce many catalogues (more or less complete) of thousands binaries. Here we present a non-exhaustive list of photometric and spectroscopic catalogues of binaries, and finally focus on the spectroscopically unresolved cases.    

\subsection{Photometric surveys}
\label{sec:eb}
Modelling the light curve of an EB provides the sum of the fractional radii relative to the semi-major axis, $(R_1+R_2)/a$, the ratio of effective temperatures, the period, the eccentricity, and the orbital inclination. Since the nineties, several ground- and space-based massive photometric surveys have monitored a large part of the sky.

\begin{itemize}
\item The Optical Gravitational Lens Experiment \citep[OGLE,][]{udalski1992}: since 1992, this Polish survey in the direction of the Galactic Bulge and the Magellanic Clouds has detected 425\,000 EB and 25\,000 ellipsoidal variables \citep{wyrzykowski2003, graczyk2011, soszynski2016}. The orbital period distribution ranges from 0.05~d (75 min) to over 2600 d (7~years) with the bulk of the distribution below 1~d. About 150 doubly EB, \emph{i.e.} quadruple systems, were also reported \citep{zasche2019}.
\item The \emph{Kepler} mission \citep{koch2010}: a monitoring of 150\,000 main-sequence stars with a field of view in Cygnus, Lyra and Draco constellations with 3\,000 EB \citep{kirk2016, yucel2022} showing an excess at 0.25~d for contact binaries \citep{kobulnicky2022} and a broader peak at 2-3~d, and even 100 doubly EB \citep{kostov2022}.
\item The Transiting Exoplanet Survey Satellite \citep[TESS,][]{ricker2015}: a monitoring of $\sim200\,000$ bright stars covering 85\% of the sky with 5\,000 EB \citep{prsa2022, howard2022} peaking at 0.25~d for contact binaries and around 3~d for the other, and 15\,000 ellipsoidal candidates with $P\ <5$~d \citep{green2023}.
\item The All-Sky Automated Survey for Supernovae \citep[ASAS-SN,][]{kochanek2017}: this survey aims at observing 1 million variable sources over the sky down to magnitude 17, and  33\,000 EB \citep{christy2022, rowan2023} in the range [0.3, 100]~d have been detected.
\item The Vista Variables in the Vi\'{a} L\'{a}ctea ESO near-IR Galactic survey \citep[VVV,][]{minniti2010}: this survey covers 1 billion Galactic stars with 33 globular clusters and 350 open clusters among which 187\,000 EB and 18\,000 contact EB have been identified using a hierarchical classifier \citep{molnar2022}.
\end{itemize} 

Such surveys have boosted the development of asteroseismology \citep{kurtz2022} including in binary stars and led to various spectacular discoveries of pulsators like heartbeat stars which are pulsating variables in eccentric orbits where resonances are induced between dynamic tides at periastron and the free oscillation modes in the stars \citep{welsh2011}; pulsating white dwarfs in EB \citep{parsons2020}; or led to the precise characterization of masses, chemical composition and ages of binaries like $\alpha$~Cen \citep[\emph{e.g.},][]{thevenin2002}. Higher-order systems in EB, like compact hierarchical triples, are reviewed by \citet{borkovits2022}. The first doubly EB were reported by \citet{zasche2022}, and many more are now uncovered \citep{kostov2022} from TESS.

\begin{figure}
\centering
\includegraphics[width=0.50\linewidth]{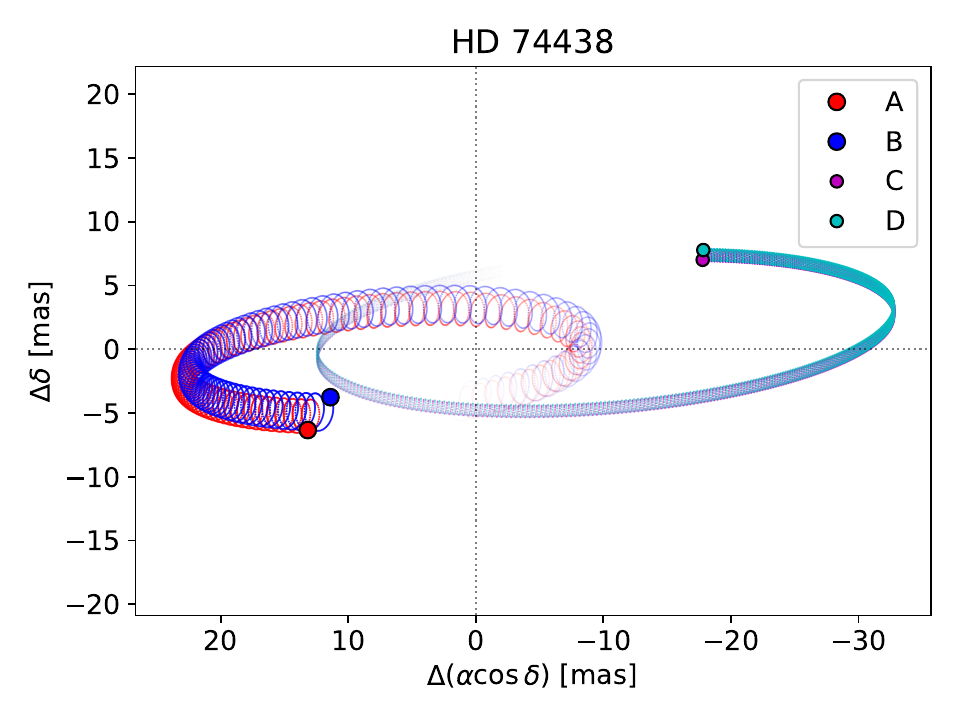}
\includegraphics[width=0.49\linewidth, clip]{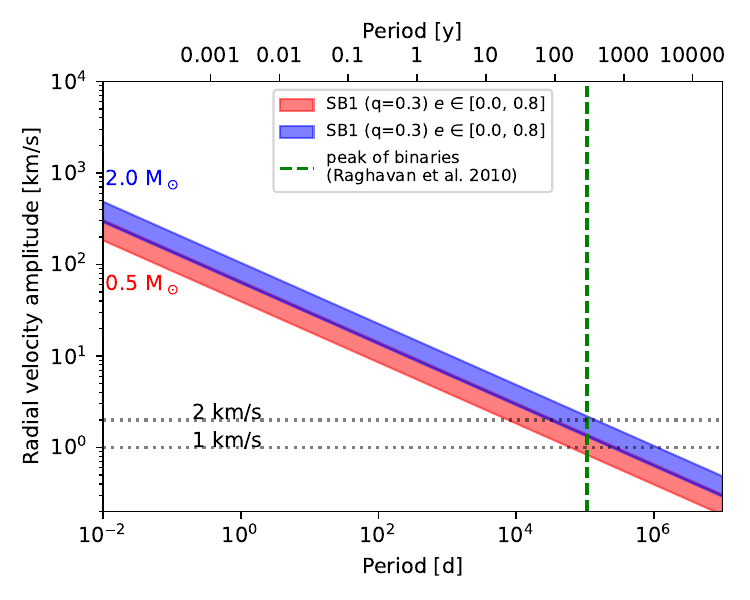}
\caption{Left: Orbital `ballet' of the four components with respect of the centre-of-mass of the spectroscopic quadruple HD~74438 from \citet{merle2022}. The four components are spatially unresolved but positions on sky can be deduced from spectroscopic and astrometric/interferometric observations. The AB, CD and AB-CD pairs have orbital periods of 20.6~d, 4.4~d and 5.7 y. We assume ascending node arguments equal to zero for the two inner orbits and a value of 274$^\circ$ for the outer orbit. For readability, the semi-major axes of the two inner orbits are  magnified by a factor of two. Right: RV semi-amplitude as a function of the orbital period for simulated SB1 with masses of the primaries at 0.5 (red) and 2.0 (blue) M$_\odot$ with a mass ratio $q=0.3$ and eccentricities in the range [0, 0.8]. The inclination on the sky is taken at 68$^\circ$.}
\label{fig:ballet}
\end{figure}

\subsection{Spectroscopic surveys}
\label{sec:sb}
Modelling the radial velocity (RV) curve of an SB2 yields the mass ratio of the components $q=M_2/M_1$, the period, the eccentricity, the RV semi-amplitudes, the centre-of-mass velocity and the projected semi-major axes. If only one spectrum is visible, it is not possible to recover the mass ratio and only the mass function is available which depends on the inclination that can be provided either by eclipses or by astrometry/interferometry. Since the new millenium, many ground-based massive spectroscopic surveys have monitored a large part of the sky:

\begin{itemize}
\item The RAdial Velocity Experiment \citep[RAVE,][]{steinmetz2006}: a survey that observed half a million stars in the Southern hemisphere at $R=7\,500$ in the Ca II triplet region. About 120 SB2 \citep{matijevic2010} and 4\,000 SB1 \citep{birko2019} were detected.
\item The Gaia-ESO Survey \citep[GES,][]{gilmore2022, randich2022} which targets 100\,000 stars in the Southern hemisphere at medium resolution ($R\sim18\,000$) and 10\,000 stars at high resolution ($R=48\,000$) among which $\sim1\,000$ SB1, SB2, SB3 \& SB4 \citep{merle2017,merle2020} and Van der Swaelmen et al. (in prep.) were identified.
\item The Apache Point Observatory Galactic Evolution Experiment \citep[APOGEE,][]{majewski2017}: this Northern hemisphere IR survey at $R\sim 22\,500$ provides a large harvest of SB among the 150\,000 stars observed including 100 SB2 \citep{fernandez2017}, 2\,500 unresolved SB2 \citep{el-badry2018b}, 20\,000 SB1 \citep{price-whelan2020}, 7\,300 SB2, 800 SB3 and 20 SB4 \citep{kounkel2021}
\item The Galactic Archaeology with HERMES \citep[GALAH,][]{de_silva2015}: targets half a million stars in the Southern sky with $R\sim28\,000$, currently 13\,000 SB2 \citep{traven2020} have been detected.
\item The Large sky Area Multi-Object fibre Spectroscopic Telescope \citep[LAMOST,][]{zhao2012}: this Northern survey is, with the 10th release, approaching 10 millions stars with low and medium resolution. 256\,000 SB1 or variable candidates \citep{qian2019}, 2\,200 SB2 \citep{zhang2022}, 3\,100 SB2, 130 SB3 \citep{li2021}, 2\,500 \citep{kovalev2022} have been identified.
\end{itemize}

The Survey of Surveys \citep[SoS,][]{tsantaki2022} have recently combined the above surveys and \emph{Gaia} DR2, providing the largest ever compilation of homogenised radial velocities for the Milky Way, which included about 10\% of binaries over the 11 millions stars (\texttt{flag\_binary}~$=1$).
Despite these impressive growing numbers, we stress that only a small fraction of a few hundreds to thousands of these detections have orbital solutions yet. Most of them are SB candidates which will require follow-up monitoring to complete their orbital phases. 

An illustrative example is provided by a spectroscopic quadruple (SB4, see left panel of Fig.~\ref{fig:ballet}), first identified in  GES \citep{merle2017} for which monitoring observations have been obtained with high resolution spectrographs (HRS at the Southern African Large telescope and HERCULES at University of Canterbury Mount-John Observatory). These monitoring allows to characterise the 2+2 architecture and the orbital parameters of the two inner and outer orbits \citep{merle2022} except the mutual inclinations for which interferometric observations would be required.

\subsection{The unresolved cases}

Unresolved SB are SB with two or more components contributing to the stellar flux of the multiple system but where the components are not resolvable by the spectrograph, either because the RV of the components are, by chance, overimposed at the epoch of the observation, or because the orbital period is long enough to have RV amplitudes smaller or of the order of the resolution of the spectrograph. For a spectrograph with a resolving power of 20\,000, this corresponds to an RV amplitude smaller or equal to the instrumental broadening of 15~km~s$^{-1}$. That is why higher the resolving power of the spectrograph is, longer the periods it can detect. According to the log-normal distribution of late-type stars binaries (\emph{e.g.}, \citealt{raghavan2010}) the main reservoir of binaries peak at a period of about 270 y. This translates to  RV semi-amplitudes between 1 and 2 km~s$^{-1}$ for a binary made of late-type primaries in the range [0.5, 2.0]~M$_\odot$ with a mass-ratio of 0.3 (right panel of Fig.~\ref{fig:ballet}), which would need to have a spectrograph with a resolving power of 150\,000 $-$ 300\,000, which is not yet within the capabilities of the today's massive multi-objects spectroscopic surveys.

Such unresolved binaries, despite the lack of sensitivity to the Doppler shift of the components, can nevertheless be identified and characterised. The proof-of-concept has been developed by \citet{el-badry2018a} and tested on APOGEE, LAMOST and GALAH-like spectra using binary against single star synthetic spectra and showing that temperature biases as large as 300~K can be obtain when a SB is treated as a single star. Using a supervised machine-learning machinery called \emph{The Payne} \citep{ting2019} and adapted for binaries, \citet{el-badry2018b} successfully identified more than 3\,000 unresolved SB in APOGEE. 

\begin{figure}
\centering
\includegraphics[width=\linewidth]{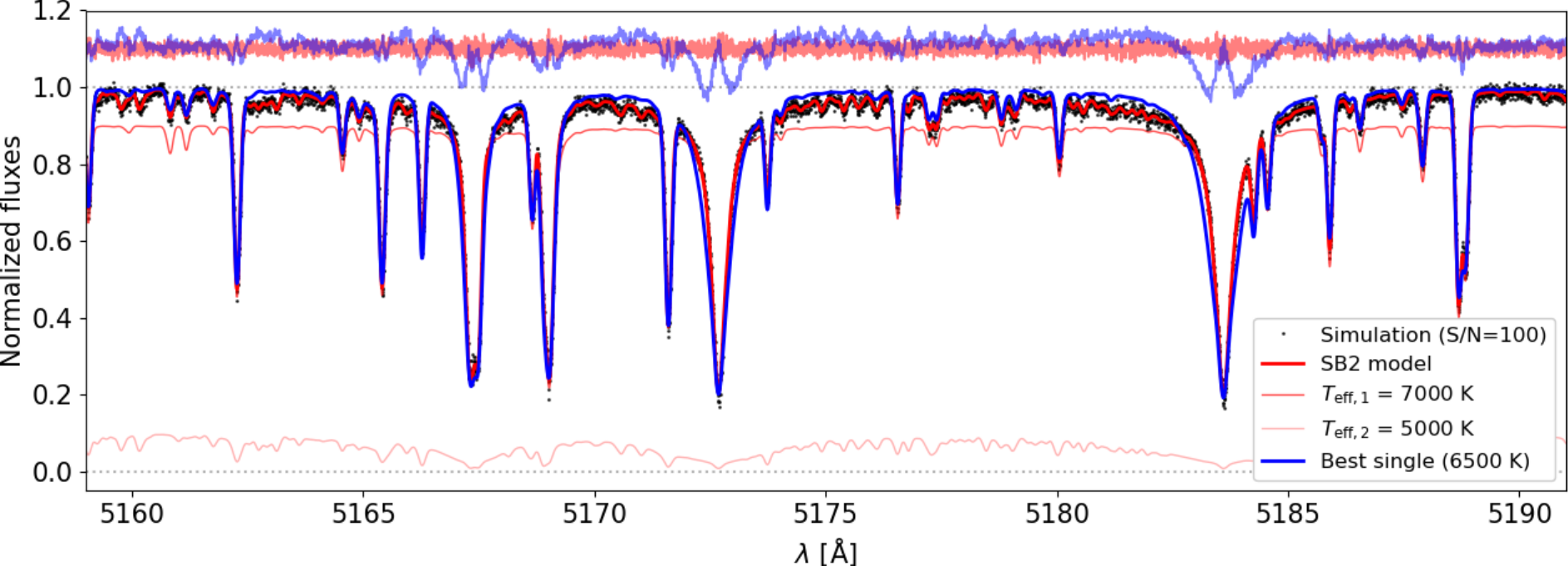}
\caption{Simulation at a resolving power of 47\,000 of an unresolved SB2 around the Mg~I~b triplet with $S/N=100$ (black dots) fitted with (i) a single star model in blue, (ii) a binary star model in red. The contributions of the two  components are also shown in light red. The residuals are shown around 1.1. The residuals of the single model show significant dispersion compared to the binary model suggesting that such unresolved SB2 can be uncovered when the companion has a flux higher than the noise level, \emph{i.e.} a few percents in this example.}
\label{fig:usb2}
\end{figure}

While such an approach is more efficient in the infra-red where the flux ratios are larger than in the visible and in the ultra-violet, it is nowadays desirable to apply systematically this kind of method to unravel many SB hidden in their long-period orbits. An illustrative example in the visible (Fig.~\ref{fig:usb2}) is performed around the Mg~I~b triplet where the  SB2 model (red) better reproduces the wings of the Mg lines compared to the best single star model (in blue). The method will be heavily sensitive to the signal-to-noise ratio because the secondary will be detectable only if its contribution to the total flux is larger than the noise.   


\section{The \emph{Gaia} revolution}
The \emph{Gaia} ESA mission \citep{perryman2001} is the first massive survey that simultaneously combined astrometric, photometric and spectroscopic detection techniques. The \emph{Gaia} DR3 \citep{gaia2022a} indeed provides a new quantitive leap in the study of binaries and multiples with the publication of the Non-Single Star catalogue \citep[NSS,][]{gaia2022b} which contains the largest homogeneous sample of about 800\,000 binaries including 87\,000 eclipsing ones, 277\,000 spectroscopic ones and 508\,000 astrometric ones. Complete orbital parameters are available for a large subset of them (\texttt{nss\_two\_body\_orbit}), while partial and tentative solutions complete the sample (\texttt{nss\_acceleration\_astro} and \texttt{nss\_non\_linear\_spectro}). We also note that, independently of the NSS catalogue, the variability catalogue  \texttt{vari\_eclipsing\_binary} provides more than 2 millions EB with detections and partial orbital solutions. By combining different methods, the DR3 also provides the estimation of masses for 195\,000 primaries and 29\,000 secondaries\footnote{\url{https://doi.org/10.17876/gaia/dr.3/56}}.

\subsection{The Hertzsprung-Russell diagram}
The Hertzsprung-Russell diagram already benefited from the precision, accuracy, and homogeneity of both astrometry and photometry from \emph{Gaia} DR2 \citep{gaia2018_brown,gaia2019_eyer} allowing detailed studies of the various Milky Way stellar populations and stellar evolutionary phases. In particular, the 100 pc sample clearly reveals a spread above the sequence of normal stars \citep[][their Fig.~8]{gaia2018} and white dwarfs \citep[][their Fig.~1]{rebassa-mansergas2021} that reaches $-0.75$ mag corresponding to a sequence of equal-mass binaries which are photometrically unresolved. Such unresolved populations of binaries are well illustrated in the \emph{Gaia} Catalogue of Nearby Stars  \citep[GCNS,][their Fig. 32]{gaia2021_smart}. For the Hyades, the closest open cluster located at 47~pc, the authors estimate a binary fraction of 34\% for stars with masses in the range [0.2, 1.4]~M$_\odot$. At the end of stellar evolution, we can mention short-period, post-common envelope binaries including a white-dwarf and (i) a main sequence companion which are progenitors of normal cataclysmic variables (CV) as classical/dwarf novae or other polars like AM Her; or (ii) an evolved companion (evolved CV) which are progenitors of extremely low-mass WD, double white dwarfs, or AM CVn systems \citep{el-badry2021b,ren2023}.

\begin{figure}
\centering
\includegraphics[width=0.67\linewidth]{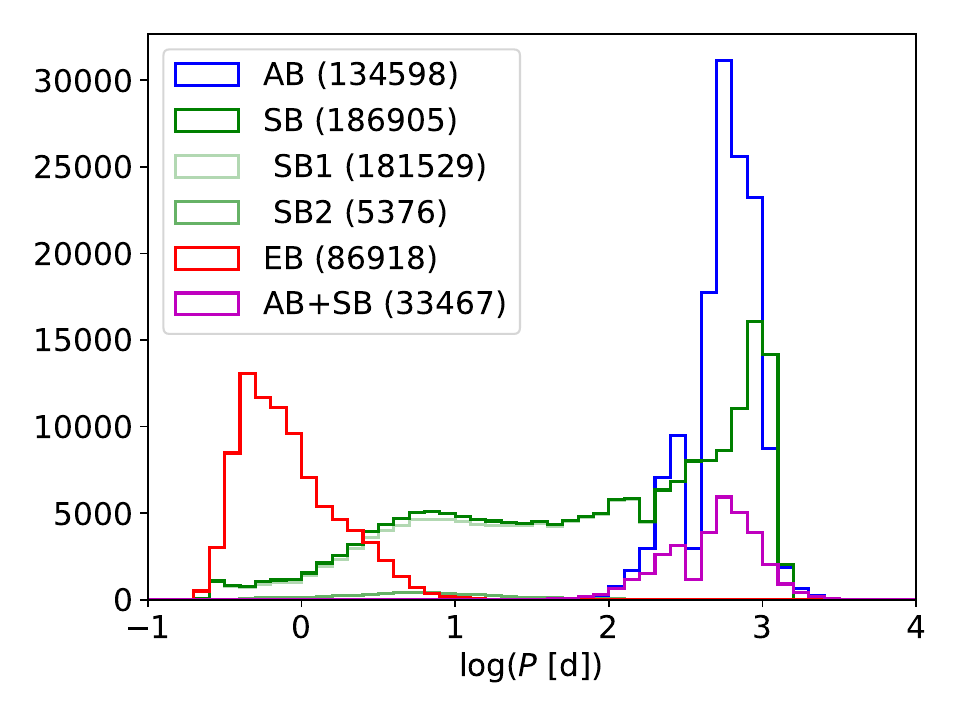}
\caption{\emph{Gaia} DR3 orbital distribution of NSS targets sorted by their astrometric (AB), spectroscopic (SB) or photometric (EB) detection method. The numbers in each category are given. The SB cover the largest number of period decades (about 4) while EB and AB cover, respectively, the short periods (below 10~d) and the longer periods (above 100~d). Note the bump around 10~d for the SB orbital period distribution.}
\label{fig:gaia_histo_period}
\end{figure}

\subsection{The orbital period distribution}
The binaries characterised in \emph{Gaia} DR3 are those for which an orbital solution has been derived from the 34 months (2.8~years) of data collection. The SB, AB and EB reach the limiting $G$ magnitudes of $\sim13$, 16 and 19, respectively. The distribution of their orbital periods are displayed on Fig.~\ref{fig:gaia_histo_period}. EB and AB probe very different period regimes (90\% of the distribution ranging from 0.3 to 3.4~d and 207 to 1090~d, respectively). On the contrary, SB cover the widest range from 0.3 to 1452~d (1.5 to 1091~d for 90\% of the distribution) except when restricted to SB2, that have periods lower than 100~d (with 50\% of them having periods lower than 6~d). In addition, 33\,000 systems have a combined astrometric+spectroscopic solution covering the same period range as AB. 155 systems also have eclipsing+spectroscopic solutions with a magnitude range from 9 to 13. The lack of AB around $\log{P}=2.5$ (\emph{i.e.}, systems with period of one year) is related to the difficulty to decouple the orbital motion from the parallactic effect.
For EB, the peak at $\sim0.5$~d corresponds to contact binaries, when both stars fill their respective Roche lobe. The gradual drop-off at the right tail of the EB distribution is mainly due to geometrical effects that decrease the probability of the eclipses at longer periods. \citealt{rowan2023} show that the eclipse probability for main sequence stars follow a $P^{-2/3}$ law.

Concerning the 187\,000 binaries identified as SB alone, about 3\% are SB2 (5\,400 systems) and about 0.5\% (950 systems) have a circular orbit. 
The shape of the SB orbital distribution shows a bump located just below 10~d and a peak at 1000~d. The latter peak is an
observational bias because the log-normal distribution of all binaries \citep[][peaking at $\log P\sim5$, \emph{i.e.}, $\sim270$~years]{raghavan2010} is truncated by the 2.8~years timespan  of the  \emph{Gaia} DR3. The bump below 10~d is more complicated to explain and does not seem to be the result of an observational bias. Indeed, it is also clearly visible on the orbital distribution of the 4\,000 SB of \emph{The 9th Catalogue of Spectroscopic Binary Orbits} \citep[SB9,][ their Fig.~3]{pourbaix2004}, but not discussed. In the \emph{Multiple Star Catalogue} \citep[MSC,][]{tokovinin2018}, an overdensity of inner binaries in triples is also reported around 10~d but assumed to be a possible selection effect. Using LAMOST and GALAH data, \citet{bashi2022} have produced a clean selection of DR3 SB1 where the peak at 10~d is well visible in the most significant sample. 
Physical models have been developed to explain such a bump, involving early stages of star formation with disc fragmentation and migration \citep[\emph{e.g.},][]{tokovinin2020}. Some other scenarios involved effects occurring at later stages like tidal effects in binaries with main-sequence stars \citep[\emph{e.g.},][]{witte2002}, outcome of common-envelope evolution \citep[\emph{e.g.},][]{podsiadlowski_2014}, secular evolution through Kozai-Lidov cycles and tidal friction implying a distant companion \citep[\emph{e.g.},][]{fabrycky2007,bataille2018}. The latter scenario is also supported by the stellar triples found by comparing Gaia NSS astrometric acceleration and SB solutions \citep{gaia2022b}, and the ones found by cross-matching with the catalogue of wide binaries \citep{el-badry2021}, which all show a peak between 5 and 10~d (see Figs. 52, 53 and 54 in \citealt{gaia2022b}).

\subsection{Binaries including a quiet compact object}
About $\sim 350$ compact stellar remnants are identified in X-ray binaries, \emph{i.e.}, with high X-ray luminosities, where a neutron star (NS, \emph{e.g.}, Sco X-1, \citealt{sandage1966}) or a black hole (BH, \emph{e.g.}, Cyg X-1, \citealt{miller-jones2021}) accretes gas from a companion through mass-transfer processes like Roche-lobe overflow or stellar winds \citep{chaty2022}.  

The recent discoveries of binary candidates including quiescent compact companions like NS \citep{mazeh2022, escorza2023} or stellar BH \citep{shenar2022}, have open the lanes to hunt such candidates in a more systematic way (\emph{e.g.}, \citealt{mahy2022,shahaf2023}), allowing the detailed investigations of the stellar graveyard of the Milky Way and its surroundings. This happened after the \emph{BH police}\footnote{ \url{https://www.eso.org/public/news/eso2210/}} discarded several claims of such detections of binaries with radio-quiet compact objects. For instance, thanks to the infrared spectro-interferometry, \citet{frost2022} rejected for HR~6819 the model of a hierarchical triple with an inner binary harbouring a dormant BH.

Very recently, the first \emph{Gaia} binaries hosting quiet BH, and coined \emph{Gaia} BH1 and BH2, have been convincingly identified \citep{el-badry2023a,el-badry2023b}, albeit the authors still considered them as candidates. \emph{Gaia} BH1 is the first confirmed Sun-like star orbiting a black hole \citep{el-badry2023a}, and \emph{Gaia} BH2 is a red giant orbiting a black hole \citep{el-badry2023b}, which make them both the widest systems that include a BH (186~d and 1280~d) but most importantly, the two BH closest to our Solar System (0.5 and 1.2 kpc).

\section{Conclusion}
We are living in an exciting era where astronomers are not limited anymore by the quantity, the quality and the homogeneity of the data. But this should not prevent us to forget about the founding works of our elders, and in addition to the massive photometric and spectroscopic surveys presented in Sects.~\ref{sec:eb} and \ref{sec:sb} which often host their own databases, it is also important to acknowledge the huge efforts performed to maintain ancillary databases of binary stars:
\begin{itemize}
\item The Washington Double Star catalogue \citep[\href{http://www.astro.gsu.edu/wds/}{WDS},][]{mason2001} maintained by the US Naval Observatory since 1964 is the largest database of visual/astrometric/interferometric binaries and multiples including more than 155\,000 systems;   
\item The Binary star DataBase \citep[\href{http://bdb.inasan.ru/}{BDB},][]{kovaleva2015} hosted since 2008 by the Russian Academy of Science was originally developed at the Besan\c{ç}on Observatory in 1995 in collaboration with D. Pourbaix who laid the first computing foundations, provides the most comprehensive information about all types of binaries and contains about 120\,000 systems;  
\item The General Catalogue of Variable Stars \citep[\href{http://www.sai.msu.su/gcvs/gcvs/}{GCVS 5.1},][]{samus2017} hosted by the Russian Academy of Science and NASA which includes cataclysmic variables, EB systems and variable X-ray sources; 
\item The Ninth Catalogue of Spectroscopic Binary Orbits \citep[\href{https://sb9.astro.ulb.ac.be/}{SB9},][]{pourbaix2004} hosted at ULB since 2000 and founded more than a century ago by \citet{campbell1905}, contains more than 4\,000 SB (2\,800 SB1 and 1\,200 SB2) in the last update by D. Pourbaix in 2021-03-02; 
\item The Multiple Star Catalogue \citep[\href{https://www.ctio.noirlab.edu/~atokovin/stars/}{MSC},][]{tokovinin2018} hosted at NOIRLab and containing more than 5\,000 stellar triples and higher-order systems (including septuplets) in the last release (2023-02-07);
\item The Detached Eclipsing Binary Catalogue \citep[\href{https://www.astro.keele.ac.uk/jkt/debcat/}{DEBCat},][]{southworth2015} hosted at Keele University and containing more than 320 well-characterised systems. 
\end{itemize}  
Obviously, this list is far from being complete, and many additional binary catalogues can be found on the VizieR catalogue service \citep{oschsenbein2000} by querying, in the `Astronomy' field: \texttt{Binaries:eclipsing}, \texttt{Binaries:spectroscopic} and \texttt{Binaries:cataclysmic} which return 350+, 790 and 300+ catalogues, respectively (not all of them being relevant, though).

\begin{acknowledgments}
I would like to thank A. Jorissen and S. Van Eck for the careful proof-reading of this manuscript and the useful discussions we had.  
The Indian and Belgian funding agencies DST (DST/INT/Belg/P-09/2017) and BELSPO (BL/33/IN12) are acknowledged for providing  financial support to participate in the third BINA workshop.
This research has made use of NASA's Astrophysics Data System Bibliographic Services; of the SIMBAD database and the VizieR catalogue access tool provided and operated at CDS, Strasbourg, France.
This work has made use of data from the European Space Agency (ESA) mission Gaia (\url{https://www.cosmos.esa.int/gaia}), processed by the Gaia Data Processing and Analysis Consortium (DPAC, \url{https://www.cosmos.esa.int/web/gaia/dpac/consortium}). Funding for the DPAC has been provided by national institutions, in particular the institutions participating in the Gaia Multilateral Agreement.
This work has made use of python 3.x (\url{https://www.python.org}) and of the following python's modules: astropy (\url{https://www.astropy.org}), a community-developed core Python package and an ecosystem of tools and resources for astronomy; matplotlib (\url{https://matplotlib.org/}); numpy (\url{https://numpy.org/}) and scipy (\url{https://scipy.org/}).
\end{acknowledgments}

\begin{furtherinformation}

\begin{orcids}
\orcid{0000-0001-8253-1603}{Thibault}{Merle}
\end{orcids}

\begin{conflictsofinterest}
The author is granted by the BELSPO Belgian federal research program FED-tWIN under the research profile Prf-2020-033\_BISTRO.
\end{conflictsofinterest}

\end{furtherinformation}

\bibliographystyle{bullsrsl-en}

\bibliography{merle_review_revised2}

\end{document}